\documentclass[prl,amsmath,amssymb,twocolumn,showpacs,superscriptaddress]{revtex4-1}
\usepackage{graphicx}

\usepackage{color}
\usepackage{amsmath}
\usepackage{bm}

\begin{document}

\title{Huge thermoelectric effects in ferromagnet--superconductor junctions in the presence of a spin-splitting field}

\author{A.~Ozaeta}

\affiliation{
Centro de F\'{i}sica de Materiales (CFM-MPC), Centro Mixto CSIC-UPV/EHU, Manuel de Lardizabal 5, E-20018 San Sebasti\'{a}n, Spain}

\author{P.~Virtanen}

\affiliation{Low Temperature Laboratory, Aalto University, P.O. Box 15100, FI-00076 AALTO, Finland}

\author{F.S.~Bergeret}

\affiliation{
Centro de F\'{i}sica de Materiales (CFM-MPC), Centro Mixto CSIC-UPV/EHU, Manuel de Lardizabal 5, E-20018 San Sebasti\'{a}n, Spain}

\affiliation{Donostia International Physics Center (DIPC), Manuel de Lardizabal 5, E-20018 San Sebasti\'{a}n, Spain}

\affiliation{Institut f\"ur Physik, Carl von Ossietzky Universit\"at, D-26111
Oldenburg, Germany}

\author{T.T.~Heikkil\"a}

\affiliation{Low Temperature Laboratory, Aalto University, P.O. Box 15100, FI-00076 AALTO, Finland}

\newcommand{\tmpnote}[1]%
   {\begingroup{\color{blue}\it (FIXME: #1)}\endgroup}

\date{\today}

\begin{abstract}
  We show that a huge thermoelectric effect can be observed by
  contacting a superconductor whose density of states is spin-split by
  a Zeeman field with a ferromagnet with a non-zero
  polarization. The resulting thermopower exceeds $k_B/e$ by a large
  factor, and the thermoelectric figure of merit $ZT$ can far exceed
  unity, leading to heat engine efficiencies close to the Carnot limit.  We
  also show that spin-polarized currents can be generated in the superconductor by
  applying a temperature bias.
\end{abstract}


\maketitle

Thermoelectric effects, electric potentials generated by temperature
gradients and vice versa, are intensely studied because of their
possible use in converting the waste heat from various processes to
useful energy. The conversion efficiency $\eta=\dot{W}/\dot{Q}$, the
ratio of output power $\dot{W}$ to the rate of thermal energy consumed
$\dot{Q}$, in thermoelectric devices however typically falls short of
the theoretical Carnot limit and is low compared to other heat engines,
which has motivated an extensive search for better
materials. \cite{shakouri2011}

In electronic conductors a major contributor to thermoelectricity is
breaking of the symmetry between positive and negative-energy charge
carriers (electrons and holes, respectively)
\cite{ashcroftmermin}. Within Sommerfeld expansion, this is described by
the Mott relation \cite{cutler69}, which predicts thermoelectric
effects of the order $\sim{}k_B T/E_0$, where $T$ is the temperature
and $E_0$ a microscopic energy scale describing the energy dependence
in the transport. This is usually a large atomic energy scale (in
metals, the Fermi energy), so that $E_0\gg{}k_BT$ even at room
temperature and these effects are often weak. Larger electron-hole
asymmetries are however attainable in semiconductors, as the chemical
potential can be tuned close to the band edges, where the density of
states varies rapidly. \cite{shakouri2011,mahan1989-fmt}

The situation in superconductors is superficially similar to
semiconductors. The quasiparticle transport is naturally strongly
energy dependent due to the presence of the  energy gap $\Delta$,
which can be significantly smaller than atomic energy scales. However,
the chemical potential is not tunable in the same sense as in
semiconductors, as charge neutrality dictates that electron-hole
symmetry around the chemical potential is preserved. This implies that
the thermoelectric effects in superconductors are often even weaker
than in the corresponding normal state, in addition to being masked
by supercurrents \cite{ginzburg44,galperin02}.

We show in this Letter that this problem can be overcome in a
conventional superconductor by applying a spin-splitting field $h$. It
shifts the energies of electrons with parallel and antiparallel spin
orientations to opposite directions.  \cite{Tedrow1971} This breaks the electron-hole
symmetry for each spin separately, but conserves charge neutrality, as
the total density of states remains electron-hole symmetric. In this
situation, thermoelectric effects can be obtained by coupling the
superconductor to a spin-polarized system.


\begin{figure}[h]
  \centering
  \includegraphics[width=\columnwidth]{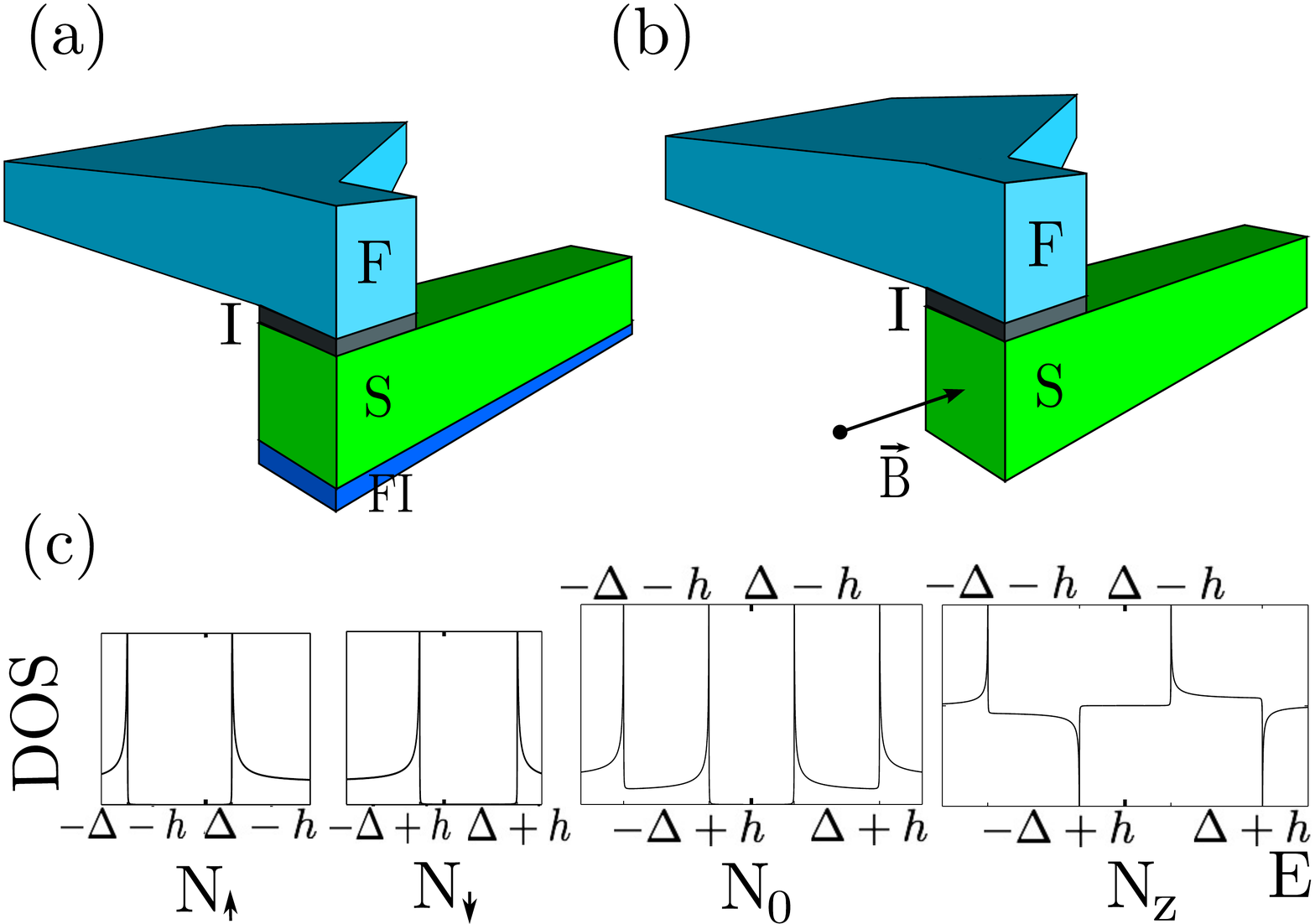}
  \caption{(Color online)Top: Schematic systems studied in this work. In both of them
  a ferromagnet (F) is coupled via a tunneling contact to a
  superconductor (S), whose tunneling density of states is modified by
  an exchange field. In (a) the exchange field is induced by the
  proximity of a ferromagnetic insulator (FI), whereas in (b) it is
  induced by the Zeeman energy due to an applied magnetic field
  $\vec{B}$ parallel with the easy axis of the ferromagnet. Bottom:
  Tunneling densities of states for spin $\uparrow/\downarrow$,
  averaged over spin ($N_0$), and the difference of them ($N_z$)
  obtained for an exchange field $h=\Delta/2$.}
\label{fig:system}
\end{figure}

We propose that this effect can be realized in structures such as
shown schematically in Fig.~\ref{fig:system}: There, a ferromagnet
with a relatively large spin polarization is connected to a
superconductor via a tunnel contact. Moreover, we assume the presence
of a finite exchange field $h$ inside the superconductor. Such an
exchange field can result from a Zeeman effect due to an applied
magnetic field (Fig.~\ref{fig:system}b) \cite{Tedrow1971}, or from a
magnetic proximity effect with either a ferromagnetic insulator
\cite{Moodera1990,Moodera2013,Tokuyasu88} or with a thin ferromagnetic metallic
layer \cite{Bergeret2001, giazotto07} placed directly below the
superconductor (Fig.~\ref{fig:system}a). For simplicity, we assume
this exchange field to be collinear with the magnetization inside the
ferromagnet.

A standard tunneling Hamiltonian calculation yields for spin-$\sigma$ electrons 
from the ferromagnet the charge and heat currents
\begin{subequations}
\begin{align}
  I_\sigma &= \frac{G_\sigma}{e} \int_{-\infty}^\infty dE N_\sigma(E)[f_F(E)-f_S(E)]
  \,,
  \\
  \dot Q_\sigma &= \frac{G_\sigma}{e^2} \int_{-\infty}^\infty dE (E-\mu_F) N_\sigma(E)[f_F(E)-f_S(E)]
  \,.
\end{align}
\end{subequations}
Here $N_{\uparrow/\downarrow}(E) =N_S(E \pm h) $ is the tunneling
density of states (DOS) for spin $\uparrow/\downarrow$ particles divided by
the normal-state density of states at Fermi energy, \cite{Tedrow1971}
$N_S(E)=|E|/\sqrt{E^2-\Delta^2}\theta(|E|-\Delta)$ is the BCS DOS,
$G_\sigma$ is the conductance through the junction for spin $\sigma$
particles in the normal state, and $f_{F/S}(E)$ are the (Fermi)
distribution functions of electrons inside the ferromagnet and the
superconductor, respectively. We disregard the energy dependence of
the density of states inside the ferromagnet as well as the tiny
electron-hole asymmetry possibly existing in the
superconductor. Moreover, we fix the electrochemical potential of the
superconductor to zero and describe the applied voltage via the
potential $\mu_F=-eV$ in the ferromagnet.

The spin-dependent densities of states $N_\sigma(E)$ are plotted in
Fig.~\ref{fig:system}c in the presence of a non-zero exchange
field. We can see that they break the symmetry with respect to
positive and negative energies for each spin. This symmetry breaking
allows for the creation of a large spin-resolved thermoelectric
effect, which can be converted to a spin-averaged effect via the spin
filtering provided by the polarization $P \equiv
(G_\uparrow-G_\downarrow)/(G_\uparrow+G_\downarrow)$. This can be seen
better by introducing the charge and spin currents
$I=I_\uparrow+I_\downarrow$ and $I_S=I_\uparrow-I_\downarrow$ as well
as the heat and spin heat currents $\dot Q=\dot Q_\uparrow+\dot
Q_\downarrow$ and $\dot Q_S=\dot Q_\uparrow-\dot Q_\downarrow$ along
with $N_0 \equiv (N_\uparrow+N_\downarrow)/2$, $N_z \equiv
N_\uparrow-N_\downarrow$,
\begin{subequations}
\label{eq:currents}
\begin{align}
I&=\frac{G_T}{e}\int_{-\infty}^\infty dE \left[N_0 +\frac{P N_z}{2}\right]\left[f_F-f_S\right]
\,,
\\
I_S&=\frac{G_T}{e}\int_{-\infty}^\infty dE \left[P N_0 +\frac{N_z}{2}\right] \left[f_F-f_S\right]
\,,
\\
\dot Q&=\frac{G_T}{e^2}\int_{-\infty}^\infty dE (E-\mu_F)\left[N_0 +\frac{P N_z}{2}\right] \left[f_F-f_S\right]
\,,
\\
\dot Q_S&=\frac{G_T}{e^2}\int_{-\infty}^\infty dE (E-\mu_F) \left[P N_0 +\frac{N_z}{2}\right] \left[f_F-f_S\right]
\,.
\end{align}
\end{subequations}
Here $G_T=G_\uparrow+G_\downarrow$ is the conductance of the tunnel
junction that would be measured in the absence of
superconductivity. The average density of states $N_0(E)$ is symmetric
and the difference $N_z(E)$ antisymmetric with respect to $E=0$ as
shown in Fig.~\ref{fig:system}c.  This means that they will pick up a
different symmetry component of the distribution function difference
in Eqs.~\eqref{eq:currents} and eventually lead to a thermoelectric
effect.

In order to grasp the size of the thermoelectric effects we assume
either a small voltage $V$ or a small temperature difference $\Delta
T/T=2(T_L-T_R)/(T_L+T_R)$ across the junctions and find the currents in
Eqs.~\eqref{eq:currents} up to linear order in $V$ and $\Delta
T/T$. They can be written in a compact way, for the charge and heat
currents
\begin{equation}
  \begin{pmatrix} I \\ \dot Q  \end{pmatrix} = \begin{pmatrix} G & P \alpha  \\ P \alpha & G_{th} T  \end{pmatrix} \begin{pmatrix} V \\ \Delta T/T \end{pmatrix}\; ,\label{response1}
\end{equation}
and for the spin and spin heat currents
\begin{equation}
  \begin{pmatrix}  I_S \\ \dot Q_S \end{pmatrix} = \begin{pmatrix} P G & \alpha \\ \alpha & P G_{th} T \end{pmatrix} \begin{pmatrix} V \\ \Delta T/T \end{pmatrix}\; .\label{response2}
\end{equation}
\begin{figure}[h]
  \centering
  \includegraphics[width=\columnwidth]{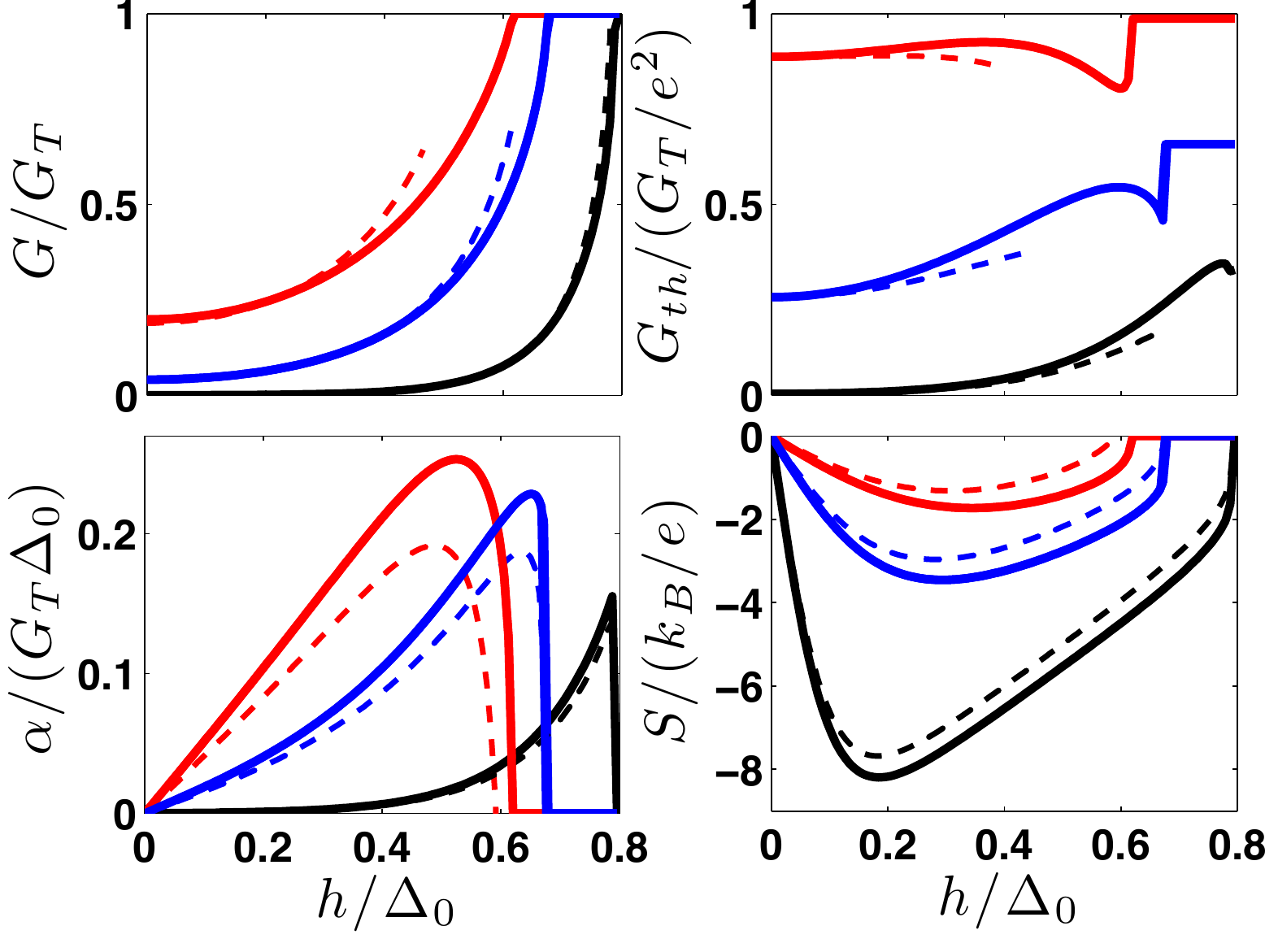}
  \caption{(Color online) Thermoelectric coefficients vs. exchange
    field $h$ at $k_B T/\Delta_0=0.1$ (black), 0.2 (blue) and 0.3
    (red). From top left to bottom right: conductance, heat
    conductance, thermoelectric coefficient, and thermopower. The
    solid lines are numerical integrals of Eqs.~\eqref{eq:numcoefs},
    the dashed lines are the approximations in
    Eqs.~(\ref{eq:coefs},8). The curves have been calculated for $\Gamma
    = 10^{-6}\Delta_0$. $\Delta_0$ is the superconducting order parameter at $T=0$ and $h=0$. }
  \label{fig:coefs}
\end{figure}
These response matrices  are expressed in terms of three coefficients,
\begin{subequations}
  \label{eq:numcoefs}
  \begin{align}
    G&=G_T \int_{-\infty}^\infty dE \frac{N_0(E)}{4 k_B T \cosh^2\left(\frac{E}{2k_B T}\right)}
    \,,
    \\
    G_{th}&=\frac{G_T}{e^2} \int_{-\infty}^\infty dE \frac{E^2 N_0(E)}{4 k_B T^2 \cosh^2\left(\frac{E}{2k_B T}\right)}
    \,,
    \\
    \alpha&=\frac{G_T}{2e} \int_{-\infty}^\infty dE \frac{E N_z(E)}{4 k_B T \cosh^2\left(\frac{E}{2k_B T}\right)}
    \,.
\end{align}
\end{subequations}
Besides the thermoelectric effect that is detailed below, we can
already draw some important conclusions based on
Eqs. (\ref{response1}-\ref{eq:numcoefs}): (i) The matrices in
Eqs. (\ref{response1}-\ref{response2}) obey the Onsager reciprocal
relations \cite{onsager31,jacquod12,machon13}, which for a generic
thermoelectric response matrix $L$ describing response in a magnetic
field $\vec{B}$ for magnetization $\vec{m}$ reads
$L(\vec{B},\vec{m})=L^T(-\vec{B},-\vec{m})$.  
Moreover, the coefficients satisfy a thermodynamic stability condition
$\alpha^2/(TGG_{\rm th})\le1$, due to Cauchy-Schwartz inequality.
(ii) The thermoelectric
effects vanish when $N_z=0$, i.e., when either no
exchange field is applied ($h=0$) or when $\Delta=0$.  Since
$N_z(-h)=-N_z(h)$, inverting the exchange field changes the sign of
the thermoelectric coefficients. It is important to emphasize that in
order to get a non-zero spin-averaged thermoelectric effect, the spin
polarization $P$ of the interface needs to be non-vanishing.  (iii)
According to Eq. (\ref{response2}), a finite spin-polarized current
can flow if there is a temperature difference across the junction.
This effect is the longitudinal analog to the spin-Seebeck effect
observed in metallic magnets \cite{uchida2008,adachi2013}, and can here
be found in a spin-splitting field even for a zero spin polarization
$P=0$.  

The response coefficients from Eqs. (\ref{eq:numcoefs}) are plotted as a
function of exchange field $h$ in Fig.~\ref{fig:coefs}. We note that the thermoelectric coefficient $\alpha$ increases linearly
for small $h$, and reaches a maximum for $h < \Delta_0$
(here, $\Delta_0$ is the superconducting order parameter at
$T=0$ and $h=0$), and finally drops to zero when superconductivity is destroyed
by $h$. Thermal conductance $G_{th}$ has a similar non-monotonic
behavior, whereas the conductance G increases monotonically toward
its normal-state value $G_T$. In the low
temperature limit $k_B T \ll \Delta-|h|$, the coefficients can be
approximated by
\begin{subequations}
\label{eq:coefs}
\begin{align}
G&\approx G_T\sqrt{2\pi\tilde\Delta} \cosh(\tilde h)e^{-\tilde\Delta}
\,,
\\
 G_{\rm th}&\approx \frac{k_B G_T\Delta}{e^2}\sqrt{\frac{\pi}{2\tilde\Delta}}e^{-\tilde\Delta}\left[e^{\tilde h}(\tilde\Delta-\tilde h)^2+e^{-\tilde h}(\tilde\Delta+\tilde h)^2 \right]
\,,
\\
 \alpha &\approx \frac{G_T }{e}\sqrt{2\pi\tilde\Delta}e^{-\tilde\Delta}\left[\Delta\sinh(\tilde h)-h\cosh(\tilde h)\right]
\,,
 \end{align}
\end{subequations}
where $\tilde \Delta=\Delta/(k_B T)$ and $\tilde h=h/(k_B T)$. For
$h=0$, the expressions reduce to the standard results for the NIS
charge and heat conductance $G$ and $G_{\rm th}$, \cite{Nahum94,Leivo96}
whereas $\alpha$ vanishes.

Instead of the thermally induced current, the typical thermoelectric
observable is the thermopower or the Seebeck coefficient
$S=-P\alpha/(G T)$, defined as the voltage $V$ observed due to a
temperature difference $\Delta T$ after opening the circuit such that
$I=0$. It can be obtained from Eqs. \eqref{eq:numcoefs}. The Seebeck
coefficient for our FIS junction is plotted in the lower right panel
of Fig.~\ref{fig:coefs}. The qualitative behavior is close to that of
$\alpha$, but it is quantitatively changed by the $h$-dependence of
$G$.  

In the low temperature limit, $S$ can be obtained from
Eqs. \eqref{eq:coefs}, $S\approx -\frac{P \Delta}{e
T}[\tanh(\tilde{h})-h/\Delta]$.  Thus, for low temperatures the
thermopower is maximized for $h=k_B T {\rm arcosh}(\sqrt{\tilde
  \Delta})$, where
\begin{equation}
\label{eq:thermopower1}
S_{\rm max} \approx -\frac{k_B}{e} P\left[\frac{\Delta}{k_B T} -{\rm arcosh}\left(\sqrt{\frac{\Delta}{k_B T}}\right)\right]
\,,
\end{equation}
It can hence greatly exceed $k_B/e$ and seems to diverge towards low
temperatures as $1/T$. In practice this divergence is cut off by
additional contributions beyond the standard BCS tunnel formula. These
are often described via the phenomenological ``broadening'' parameter
$\Gamma$ \cite{pekola04}. Practical reasons for the occurrence of an
effectively non-zero $\Gamma$ are due to the fluctuations in the
electromagnetic environment \cite{pekola10}, the presence of Andreev
reflection \cite{rajauria08,laakso12}, or the inverse proximity effect
from the ferromagnet \cite{sillanpaa01,kauppilaup13}. The main effect
of the broadening parameter for the thermopower is to induce a finite
density of states inside the gap that in turn leads to a correction of
the charge conductance \eqref{eq:coefs} of the order $\delta G =
\frac{\Gamma}{\Delta} G_T$ (valid for $\Gamma \ll k_B T \ll
\Delta$). The corrections for the other coefficients are less
relevant. Within this limit  we get for the thermopower
\begin{align}
  S&=P\frac{\Delta }{e T}\frac{h \cosh\left(\frac{h}{k_B
        T}\right)-\Delta\sinh\left(\frac{h}{k_B T}\right)}{\Gamma
    e^{\Delta/(k_B T)}\sqrt{\frac{k_B
        T}{2\pi\Delta}}+\Delta\cosh\left(\frac{h}{k_B T}\right)}.
  \label{eq:thermopower2}
\end{align}
The result for $S$ is shown in the lower right panel of Fig.~\ref{fig:coefs}.

\begin{figure}[h]
  \centering
  \includegraphics[width=\columnwidth]{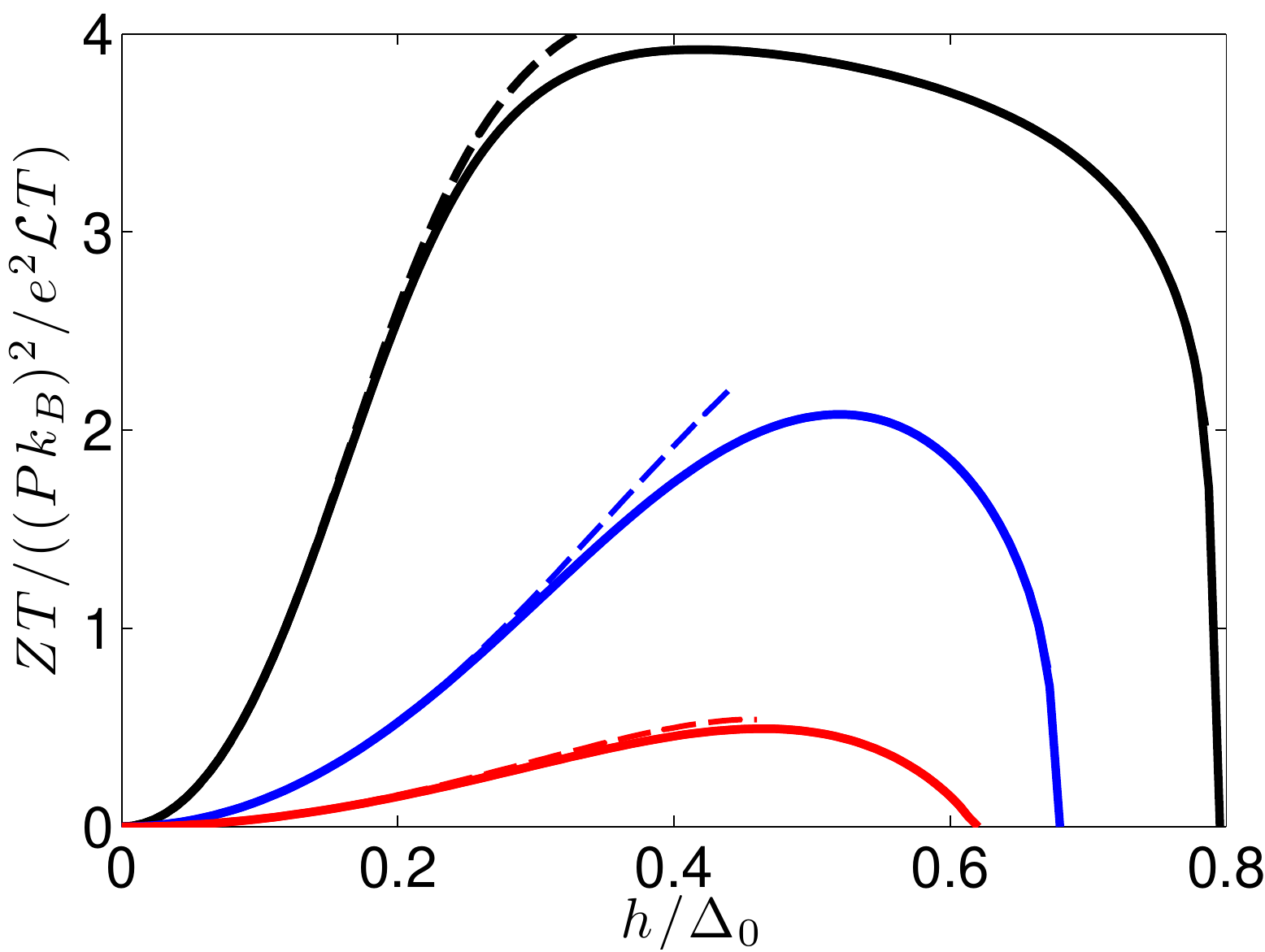}
  \caption{(Color online) Figure of merit ZT as a function of
    exchange field at $k_B T/\Delta_0=0.1$ (black), 0.2 (blue) and 0.3
    (red) and $P=0.9$. The solid line is the exact result and the dashed line the result obtained from Eq. \eqref{eq:zt}.}
  \label{fig:ZT}
\end{figure}

The power conversion ability of thermoelectric devices is usually
characterized by a dimensionless figure of merit $ZT$, which can here
be related to the junction parameters by $ZT=S^2 G T/\tilde{G}_{\rm
  th}$, where $\tilde G_{\rm th}$ is the thermal conductance at zero
current. \cite{foot} At
linear response, $\Delta T\ll{}T$, this determines the efficiency at
maximum output power, $\eta=\eta_{CA}ZT/(ZT+2)$, where
$\eta_{CA}=1-\sqrt{T_{\rm cold}/T_{\rm hot}}$ is the Curzon-Ahlborn
efficiency. \cite{curzon1975} Best known thermoelectric bulk materials
have $ZT\lesssim{}2$, but better efficiencies are achievable in
nanostructures.  \cite{shakouri2011}

Assuming that the thermal conductance is dominated by the electronic
contribution, we find at $k_B T \ll \Delta - |h|$
\begin{equation}
  ZT
  =
  \frac{
    P^2
  }{
    1 - P^2
    +
    \frac{
      \Delta^2
    }{
      \bigl[h\cosh\bigl(\frac{h}{k_BT}\bigr) - \Delta\sinh\bigl(\frac{h}{k_BT}\bigr)\bigr]^2
    }
  }
  \,,
\label{eq:zt}
\end{equation}
which is shown and compared to numerical results in Fig.~\ref{fig:ZT}.
For $k_B T \ll h$, we find $ZT=P^2/(1-P^2)$. For $P \rightarrow 1$
(half-metal injector), $ZT$ approaches infinity, and the efficiency
approaches theoretical upper bounds. From a practical point of view 
the main challenge in achieving large
values for $ZT$ is the fabrication of barriers with large spin-polarization $P$. 

\begin{figure}
  \includegraphics[width=\columnwidth]{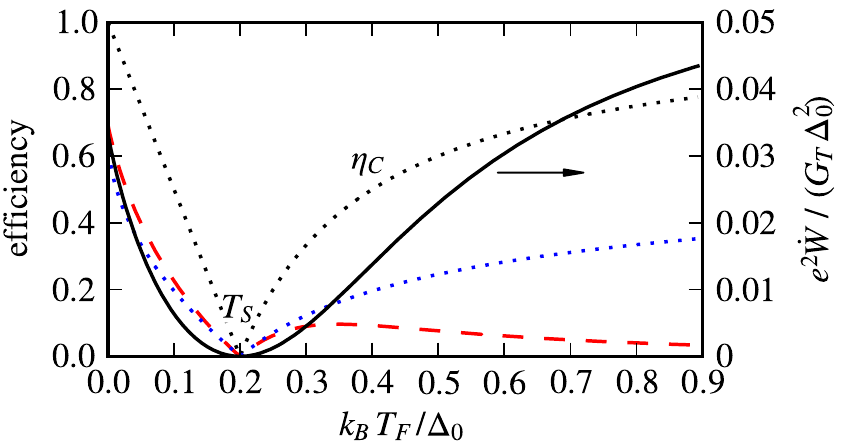}
  \caption{(Color online) Maximum power $\dot{W}=\max_V[-IV]$ generated by the FIS junction
   from a temperature difference $T_F-T_S$  (solid), 
   and the corresponding heat engine efficiency $\eta$ (dashed).
   We fix $P=1.0$, $k_BT_S=0.2\Delta_0$, $h=0.6\Delta_0$, and
   $\Gamma=10^{-5}\Delta_0$. The linear-response result $\eta=\eta_{CA}
   ZT/(ZT+2)$ for $ZT=4.04$ and the Carnot efficiency $\eta_C=1-T_{\rm
   cold}/T_{\rm hot}$ are also shown (dotted).
  }
  \label{fig:nonlinear}
\end{figure}

Let us characterize the efficiency at larger temperature
differences. Figure~\ref{fig:nonlinear} shows the maximum extractable
power as a function of the temperature difference, together with the
conversion efficiency $\eta$. For a $1\,\mathrm{k\Omega}$ tunnel
junction to aluminum, the maximum power in this figure corresponds to
$\dot{W}\approx 1.5 \,\mathrm{pW}$. The efficiency can be rather high,
$\eta=0.7$, also when the extracted power is large.

Superconductors are known to support certain thermoelectric effects
partly related to those discussed in this work. First, magnetic
impurities in superconductors can break the electron-hole symmetry and
lead to thermoelectric effects. \cite{kalenkov2012} Second, the
cooling effect found in NIS junctions in the nonlinear regime is somewhat
similar to the effect described here, if one substitutes the exchange
field with a finite voltage $V \approx \Delta/e$
\cite{giazotto06,muhonen12}. Indeed, the extracted power found above
is comparable to the maximum cooling power of a NIS junction.  NIS
junctions, however, cannot be used for power conversion, as their
cooling power $\dot{Q}_{NIS}$ is a symmetric function of the bias
voltage. The effect of ferromagnetism on NIS cooling was also
discussed earlier, \cite{giazotto02,ozaeta2012,kawabata2013} but in those works the
exchange field was introduced in order to suppress the Joule heating
due to the Andreev current and did not affect the density of the
states of the superconductor. According to our results the induced exchange field in 
the superconductor may  lead to a larger cooling  efficiency as in  NIS junctions.

The results described above are obtained by assuming that the electron
charge, spin and energy relax immediately after tunneling. This
assumption can be lifted by considering the non-equilibrium state
formed inside the ferromagnetic or the superconducting wire due to the
biasing.  This can be described by generalizing the quasiclassical
Green's function approach in Ref.~\cite{morten04} to the case of a
superconductor in a spin-splitting field, and describing the effect of
the finite spin polarization inside the ferromagnet via an effective
boundary condition derived in Ref.~\cite{bergeret12}.  We have verified
that the effects described above are qualitatively not affected by
such corrections.  The details of this approach will be published
elsewhere.

We also note that in the geometry of Fig.~\ref{fig:system}(b), where the Zeeman field is
induced by a magnetic field, the orbital effect of the magnetic field
will also influence the form of the density of states and for large
fields it will eventually lead to a destruction of
superconductivity. For simplicity, we have disregarded this effect in
the above calculation. In practice, to minimize this effect, the
magnetic field should be applied preferably in the longitudinal
direction of the wire \cite{Meservey1970}, as depicted in
Fig.~\ref{fig:system}(b).

Summarizing, we have shown that a junction between a conventional
superconductor in the presence of an exchange field and a ferromagnet
with polarization $P$ exhibits huge thermoelectric effects. The
thermopower diverges at low temperatures in the absence of limiting
effects, yielding a figure of merit $ZT \approx P^2/(1-P^2)$
and heat engine efficiencies close to theoretical upper bounds. Moreover, 
even in the case of $P=0$ our model predicts finite spin currents in the presence of a
temperature gradient, provided there is a
spin-splitting of the density of states. These mechanisms in principle can work also in semiconductors without requiring doping which typically  deteriorates the thermoelectric effects.

The authors thank V. Golovach for useful discussions. The work of F.S.B and A. O. have been supported by the Spanish
Ministry of Economy and Competitiveness under Project
FIS2011-28851-C02-02 and T.T.H. and P.V. by the Academy of Finland,
the European Research Council (Grant No. 240362-Heattronics) and the
EU-FP 7 INFERNOS (Grant No. 308850) program. The work of A. O. have
also been supported by the CSIC and the European Social Fund under
JAE-Predoc program and the EU-FP 7 MICROKELVIN project (Grant
No. 228464). A.O. acknowledges the hospitality of O.V. Lounasmaa
Laboratory (Aalto University), during his stay in Finland.

\end{document}